\documentclass[3p,12pt]{elsarticle}

\usepackage{verbatim}

\usepackage[colorlinks=true]{hyperref}
\hypersetup{urlcolor=blue, citecolor=red}

\usepackage{graphicx}
\usepackage{epstopdf}
\usepackage{floatrow}
\usepackage{subfig}
\usepackage{multirow}
\usepackage{multicol}

 \usepackage{amsmath,amsfonts,amssymb,amsthm,latexsym}
 
 \biboptions{numbers,sort&compress}

\newcounter{mnote}
\theoremstyle{plain}

\theoremstyle{definition}

\theoremstyle{remark}

\makeatletter
\def\ps@pprintTitle{%
 \let\@oddhead\@empty
 \let\@evenhead\@empty
 \def\@oddfoot{\centerline{\thepage}}%
 \let\@evenfoot\@oddfoot}
\makeatother

\def\CO2{CO$_2$}

\journal{Journal of Computational Physics}

\begin{document}

\begin{frontmatter}

\title{Reduced modeling of porous media convection in a minimal flow unit at large Rayleigh number}

\author[GEO,ICES,applied]{Baole Wen}
\ead{wenbaole@gmail.com}
\author[applied,fphys,engr]{Gregory P. Chini\corref{cor1}}
\ead{greg.chini@unh.edu}
\cortext[cor1]{\emph{Corresponding author}:  Gregory P. Chini. University of New Hampshire, 33 Academic Way, W113 Kingsbury Hall, Durham, NH 03824, USA.
Phone: +1 603 862 2633.  Fax:  +1 603 862 1865}
\address[GEO]{Department of Geological Sciences, Jackson School of Geosciences, The University of Texas at Austin, Austin, TX 78712 USA}
\address[ICES]{Institute of Computational Engineering and Sciences, The University of Texas at Austin, Austin, TX 78712 USA}
\address[applied]{Program in Integrated Applied Mathematics, University of New Hampshire, Durham, NH 03824 USA}
\address[fphys]{Center for Fluid Physics, University of New Hampshire, Durham, NH 03824 USA}
\address[engr]{Department of Mechanical Engineering, University of New Hampshire, Durham, NH 03824 USA}



\begin{abstract}
Direct numerical simulations (DNS) indicate that at large values of the Rayleigh number ($Ra$) convection in porous media self-organizes into narrowly-spaced columnar flows, with more complex spatiotemporal features being confined to boundary layers near the top and bottom walls. In this investigation of high-$Ra$ porous media convection in a minimal flow unit, two reduced modeling strategies are proposed that exploit these specific flow characteristics. Both approaches utilize the idea of decomposition since the flow exhibits different dynamics in different regions of the domain: small-scale cellular motions generally are localized within the thermal and vorticity boundary layers near the upper and lower walls, while in the interior, the flow exhibits persistent large-scale structures and only a few low (horizontal) wavenumber Fourier modes are active. Accordingly, in the first strategy, the domain is decomposed into two near-wall regions and one interior region. Our results confirm that suppressing the interior high-wavenumber modes has negligible impact on the essential structural features and transport properties of the flow. In the second strategy, a hybrid reduced model is constructed by using Galerkin projection onto a fully \emph{a priori} eigenbasis drawn from energy stability and upper bound theory, thereby extending the model reduction strategy developed by Chini \emph{et al.} (\emph{Physica~D}, vol. 240, 2011, pp. 241--248) to large $Ra$. The results indicate that the near-wall upper-bound eigenmodes can economically represent the small-scale rolls within the exquisitely-thin thermal boundary layers.  Relative to DNS, the hybrid algorithm enables over an order-of-magnitude increase in computational efficiency with only a modest loss of accuracy.
 

\end{abstract}


\begin{keyword}
Porous media convection, convection, upper bound theory, reduced-order modeling, variational analysis

\end{keyword}

\end{frontmatter}

\section{Introduction}
\label{sec:intro}

Rayleigh--B\'enard convection in a fluid-saturated porous layer is a prime example of a spatiotemporal pattern-forming system that exhibits rich nonlinear dynamics despite its comparably simple mathematical formulation \citep{Horton1945, Lapwood1948, Schubert1982, Kimura1986, Kimura1987, Aidun1987, Graham1992, Graham1994, Otero2004, Rees2011, Hewitt2012, Hewitt2013, Hewitt2014, Wen2015JFM, Hewitt2017, WenChini2018JFM,Wen2018dispersion}. Recently, there has been renewed interest in this system owing to the potential impact of buoyancy-driven convective flows on geological carbon dioxide (\CO2) storage, which is one promising means of reducing \CO2 emissions into the atmosphere \citep{Metz2005, Ennis-King2005, Riaz2006, Kim2008, Neufeld2010, Hidalgo2012, Szulczewski2013, Fu2013, Hewitt2013shutdown, Wen2018Rayleigh, Slim2014, Sathaye2014, Tilton2014, Shi2018, Akhbari2017,Wen2018}.  In a wide horizontal porous layer uniformly heated from below and cooled from above, the basic conduction state becomes unstable via a stationary bifurcation when the Rayleigh number $Ra > 4\pi^2$ \citep{Horton1945,Lapwood1948}, and convection sets in as steady O(1) aspect-ratio rolls. As $Ra$ increases, the thermal boundary layers generated by the steady roll cells become unstable and the resulting flow exhibits a series of transitions between periodic and quasi-periodic roll motions \citep{Schubert1982, Kimura1986, Kimura1987, Aidun1987, Graham1992, Graham1994}. In a two-dimensional (2D) domain, the large-scale convective rolls are completely broken down for $Ra > 1300$, culminating in spatiotemporally chaotic dynamics \cite{Otero2004}.

The direct numerical simulations (DNS) performed by \cite{Hewitt2012,Wen2015JFM} reveal that, at sufficiently large $Ra$, thermal convection in porous media exhibits a three-layer wall-normal asymptotic structure: exquisitely thin diffusive boundary layers form adjacent to the upper and lower walls; the interior region is dominated by a nearly vertical columnar exchange flow (`mega-plumes') spanning the height of the domain; and between these regions, small proto-plumes grow from the boundaries and coalesce to drive the interior mega-plumes.  As the Rayleigh number is increased, the time-mean mega-plume spacing $L_m$ shrinks as a power-law scaling of $Ra$; e.g. $L_m \sim Ra^{-0.4}$ has been proposed by~\cite{Hewitt2012}.  Moreover, the studies described in \cite{Wen2013,Dianati_thesis,Wen2015thesis} indicate that at large $Ra$ this mean inter-plume spacing approaches the minimal flow unit, above which the Nusselt number $Nu$ becomes independent of the domain aspect ratio $L$.  Collectively, these investigations suggest that the basic physics of high-$Ra$ porous media convection can be investigated in a narrow domain in which the flow retains the three-region columnar structure but includes only a single pair of rising and descending mega-plumes, and heat is transported at the same rate as in wider domains.

The complexity of turbulent flows, even in ostensibly simple configurations like high-$Ra$ porous media convection, generally necessitates the retention of hundreds of thousands or millions of degrees of freedom to ensure adequate resolution of all spatiotemporal scales of motion.  Hence it is desirable to construct models with a reduced number of degrees of freedom that capture the essential nonlinear interactions over different spatial and temporal scales.  As suggested above, one simple approach to reducing the number of degrees of freedom in turbulent flows is to study the dynamics in a small domain in which the turbulence nevertheless can be sustained.  Of course, the use of small domains precludes long-wavelength interactions, but in many cases the fundamental features of the turbulence are still retained.  In particular, we confirm in the following sections that high-$Ra$ porous media convection in a minimal flow unit exhibits the same three-region columnar structure and heat flux as is manifest in wide domains.  Therefore, in this study, we use the minimal flow unit to explore strategies for achieving further reductions in the number of degrees of freedom in simulations of porous media convection at large $Ra$.  

One popular technique for constructing low-dimensional models involves the application of the Proper Orthogonal Decomposition (POD).  In this method, an eigenfunction basis, whose modes can be ordered in terms of decreasing average energy content, is obtained by post-processing either experimental or numerical data.  Galerkin projection of the governing partial differential equations (PDEs) onto this POD basis produces a system of ordinary differential equations (ODEs), and truncations of the resulting infinite set of ODEs yield low-dimensional models.  Although POD has been used for model reduction for various turbulent flows including buoyancy-driven convection \citep{Aubry1988, Berkooz1993, Cazemier1998, Moehlis2002, Ma2002, Smith2005, Kalb2007,Bailon-Cuba2011,Chaturantabut2011,Robinson2012}, a fundamental limitation of this approach is that extensive data sets are required from experiments or DNS \emph{before} the reduced models can be constructed and their dynamics investigated.  Moreover, although a small number of POD modes may capture most of the `energy' of the infinite-dimensional dynamics, dynamically important modes having low average energy content may be omitted in the usual ordering employed in the construction of POD models \citep{CDZD2011}.

In \cite{CDZD2011}, a new, completely \emph{a priori} low-dimensional modeling strategy was proposed.  Specifically, eigenfunctions drawn from energy stability and upper bound theory were utilized to construct low-dimensional models of low-$Ra$ porous media convection.  Unlike widely-employed Fourier and Chebyshev (i.e. \emph {a priori}) basis functions, the upper bound eigenbasis is extracted directly from the governing equations and is thereby naturally adapted to the dynamics at the given parameter values.  For instance, as demonstrated in section~\ref{sec:Strategies}, at large $Ra$ certain upper-bound eigenfunctions exhibit boundary-layer structure, which is advantageous given that porous media convection self-organizes into narrow columnar plumes with more complex spatiotemporal features confined near the heated and cooled walls.  In addition, it has been shown in \cite{Wen2015JFM} that the interior flow is a composite of a few low-wavenumber Fourier modes but is dominated by a single mode; the (wall-normal varying) Fourier amplitudes of the high horizontal-wavenumber modes are strongly localized near the upper and lower walls, where they superpose to comprise the small rolls and proto-plumes within the thermal and vorticity boundary layers.  

Inspired by this emergent spatial and spectral structure of the columnar flow at large $Ra$, two complementary strategies are proposed to reduce the degrees of freedom required in numerical simulations of porous media convection.  First, a \emph{domain decomposition method}~\citep{Boyd2000} is employed: the domain is partitioned into different regions in which different resolutions are used in conjunction with a Fourier--Chebyshev pseudospectral numerical scheme; secondly, a \emph{hybrid reduced model} is constructed: at low horizontal wavenumbers, PDEs are solved using a standard Fourier--Chebyshev pseudospectral method (with domain decomposition), while at high wavenumbers, ODEs for the time-dependent coefficients of a small number of wall eigenmodes obtained from the upper-bound analysis are solved to economically capture the dynamics within the boundary layers.

The reminder of this paper is organized as follows. In the next section, we formulate the standard mathematical model of porous media convection. In section~\ref{sec:Strategies}, the two complementary strategies for reducing the degrees of freedom in numerical simulations are described in detail.  In section~\ref{sec:Results}, computations employing these two approaches are performed at large Rayleigh number in the minimal flow unit, and the results are compared with those from resolved and under-resolved DNS.  Finally, our conclusions are given in section~\ref{sec:Conclusions}.



\section{Problem formation}
\label{sec:Problemformation}
We study buoyant Darcy flow under the Boussinesq approximation  in a 2D porous layer with dimensionless horizontal and vertical coordinates $x$ and $z$, respectively.  The evolution of the (dimensionless) velocity $\mathbf{u}(\mathbf{x},t) = (u,w)$, temperature $T(\mathbf{x},t)$ and pressure $p(\mathbf{x},t)$ is governed by the Darcy--Oberbeck--Boussinesq equations \citep{NB2013} in the infinite Darcy--Prandtl number limit:
\begin{eqnarray}
    \nabla\cdot\mathbf{u} &=& 0, \label{Continuity}\\
    \mathbf{u} + \nabla p &=& RaT{\bf e}_z, \label{Darcy} \\ 
    \partial_t{T} + \mathbf{u}\cdot\nabla{T} &=& \nabla^2{T}, \label{EnergyEQ} 
\end{eqnarray}
where ${\bf e}_z$ is a unit vector in the (upward) $z$ direction and $\nabla^2$ is the 2D Laplacian operator.  These equations are solved subject to the boundary conditions
\begin{eqnarray}\label{bc}
    T(x,0,t) = 1, \quad T(x,1,t) = 0, \quad w(x,0,t) = 0, \quad w(x,1,t) = 0
\end{eqnarray}
and $L$-periodicity of all fields in the $x$ coordinate direction.  The dynamics of this system are governed by the Rayleigh number
\begin{eqnarray}\label{Rayleighnumber}
    Ra \equiv \frac{\alpha g(T_{bot} - T_{top})KH}{\nu\kappa},
\end{eqnarray}
representing the ratio of driving to damping forces, where $\alpha$ is the thermal expansion coefficient, $g$ is the gravitational acceleration, $T_{bot} - T_{top}$ is the dimensional temperature difference across the layer, $K$ is the medium permeability, $H$ is the layer depth, $\nu$ is the kinematic viscosity and $\kappa$ is the thermal diffusivity; and the domain aspect ratio $L$, the ratio of the horizontal to the vertical dimension of the domain.  A primary quantity of interest in convection is the heat transport enhancement factor, i.e. the Nusselt number $Nu$, quantifying the strength of the convective motion:
\begin{eqnarray}\label{Nusseltnumber}
    Nu \equiv 1 + \frac{1}{L}\Bigg\langle \int wT dxdz\Bigg\rangle, 
\end{eqnarray}
where the angle brackets denote a long--time average so that, for some function $f$,
\begin{eqnarray}\label{Nusseltnumber1}
    \langle f \rangle = \lim\limits_{\widetilde{t}\rightarrow \infty} \frac{1}{\widetilde{t}}\int_0^{\widetilde{t}}f dt.
\end{eqnarray}
From the equations of motion an alternative but equivalent expression for the Nusselt number can be derived \citep{Doering1998},
\begin{eqnarray}\label{Nusseltnumber2}
    Nu = -\Bigg\langle \frac{1}{L}\int \partial_z T|_{z=0} dx\Bigg\rangle \equiv -\langle\partial_z\overline{T}|_{z=0}\rangle = \langle ||\nabla T||^2 \rangle,
\end{eqnarray}
where $\overline{f} = \frac{1}{L}\int_0^L f dx$ and $||f|| = (\frac{1}{L}\int_0^1\int_0^L |f|^2dxdz)^{1/2}$.

We introduce a stream function $\psi$ to describe the fluid velocity $(u,w) = (\psi_z, -\psi_x)$, and then equations (\ref{Darcy}) and (\ref{EnergyEQ}) become
\begin{eqnarray}
    \nabla^2{\psi} &=& -Ra\frac{\partial{\theta}}{\partial{x}},\label{psi}\\
    \frac{\partial{\theta}}{\partial{t}} + \frac{\partial{\psi}}{\partial{z}}\frac{\partial{\theta}}{\partial{x}} - \frac{\partial{\psi}}{\partial{x}}\frac{\partial{\theta}}{\partial{z}} &=& \tau'\frac{\partial{\psi}}{\partial{x}} + \nabla^2{\theta} + \tau'',\label{theta}
\end{eqnarray}
where $\theta(\mathbf{x},t) = T(\mathbf{x},t) - \tau(z)$, and $\tau(z)$ either is the conduction profile (i.e. $\tau(z) = 1-z$) or the `background' profile obtained from the upper bound analysis \citep{Doering1998, Otero2004, CDZD2011, Wen2012, Wen2013}.  The fields $\theta$ and $\psi$ satisfy $L$-periodic boundary conditions in $x$ and homogeneous Dirichlet boundary conditions at $z=0,1$.  The numerical solution of (\ref{psi}) and (\ref{theta}) is facilitated by employing the spectral expansions 
\begin{eqnarray}
    \left[ \begin{array}{c} {\theta} \\ {\psi}\end{array} \right] = \sum\limits_{n = -N/2}^{N/2} \left[ \begin{array}{c} \hat{{\theta}}_n(z,t) \\ \hat{{\psi}}_n(z,t) \end{array} \right] e^{inkx}, \label{Theta_Psi_Floquet}
\end{eqnarray}
where $N$ is the horizontal truncation mode number and $k\equiv 2\pi/L$ is the fundamental wavenumber.  To evaluate the performance of our reduced models, we compare the results with data obtained from DNS.  In our DNS, temporal discretization is achieved using the Crank--Nicolson method for the linear terms and a two-step Adams--Bashforth method for the nonlinear terms, while a Fourier--Chebyshev pseudospectral collocation method is used for spatial discretization \citep{Boyd2000}.

\section{Reduced Modeling Strategies at Large $Ra$}\label{sec:Strategies}

In this section, we describe in detail two approaches for reducing the degrees of freedom required in numerical simulations of porous media convection.  It will be evident that each of these approaches exploits the high-$Ra$ structure of the convection and that the approaches are complementary, sharing certain common attributes.  

\subsection{Domain Decomposition Method}\label{sec:DomainDecomposition}

\begin{figure}[t!]
    \centering
    {\includegraphics[height=2.7in]{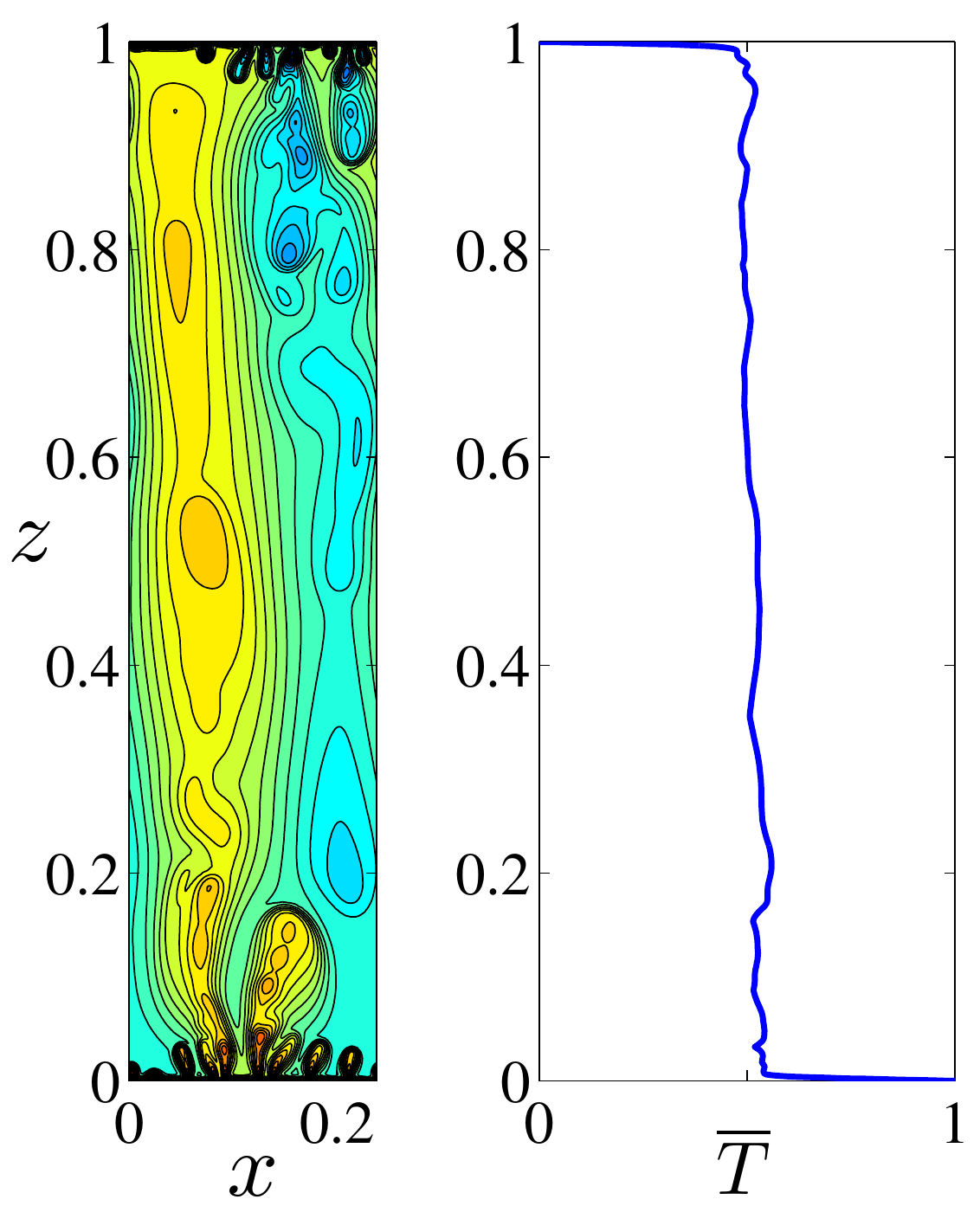}} \quad
    {\includegraphics[height=2.7in]{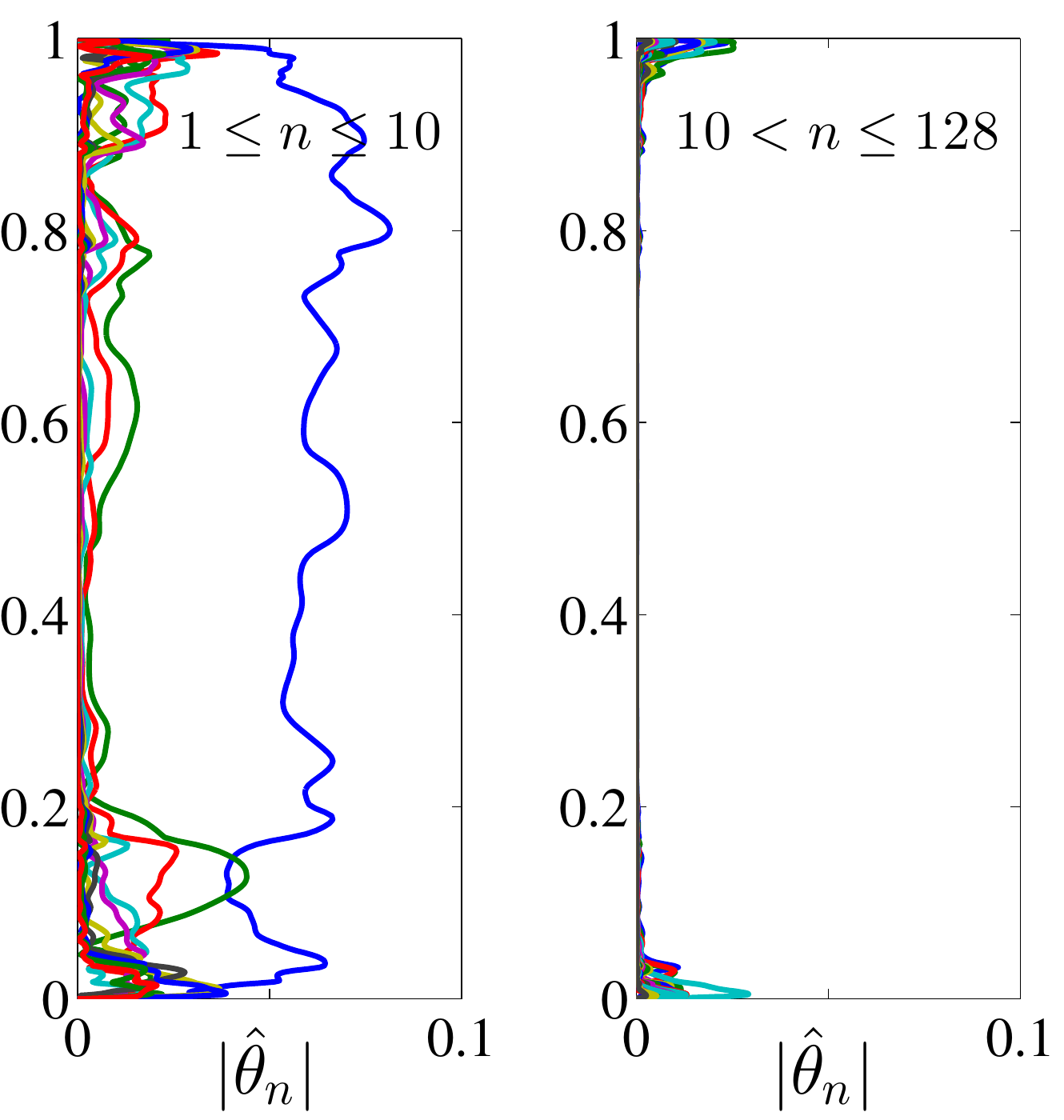}}
    \caption{A snapshot of the temperature field and the corresponding Fourier amplitudes extracted from DNS at $Ra = 20000$ and $L = 0.24$.  The numerical resolution used for this simulation is $256$ Fourier modes in $x$ and $321$ Chebyshev modes in $z$.}
    \label{Tandthetahat_MFU}
\end{figure}
\begin{figure}[h!]
    \centering
    \includegraphics[width=0.6\textwidth]{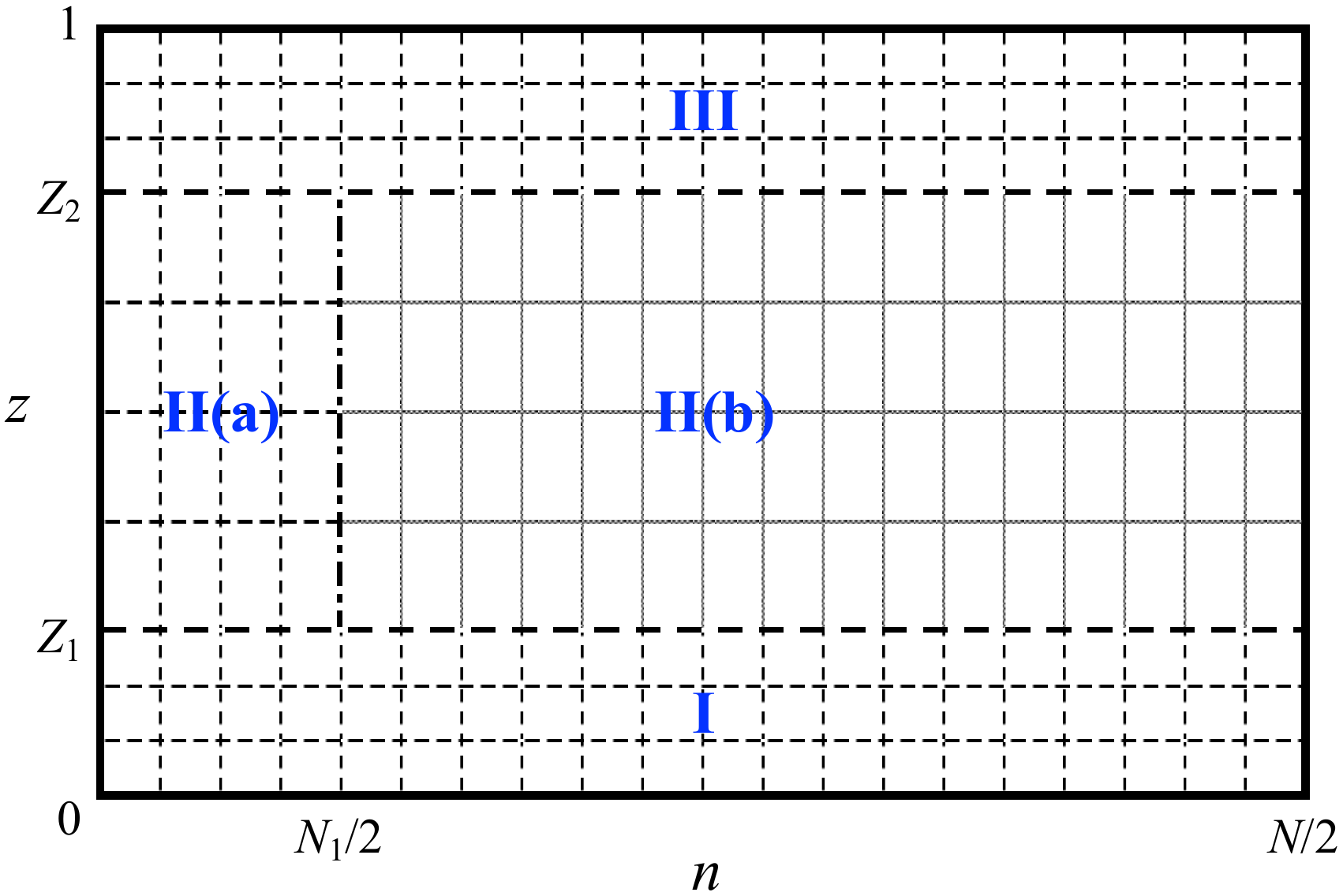}
    \caption{Schematic showing the decomposition of the domain in physical ($z$) and Fourier space at large $Ra$.  The plot only shows the positive wavenumber regime ($0 \le n \le N/2$), since the Fourier amplitudes are complex conjugates at negative wavenumbers.  Region I: $0 \le z < Z_1$; region II: $Z_1 \le z \le Z_2$ with $Z_2 = 1 - Z_1$; region III: $Z_2 < z \le 1$.  Region~II(a) includes only those low wavenumber modes satisfying $0 \le n \le N_1/2$, while region~II(b) includes the remaining high-wavenumber modes for which $N_1/2 < n \le N/2$.}
    \label{Schematic_DomDecomp}
\end{figure}

Figure~\ref{Tandthetahat_MFU} shows a snapshot of the temperature field, the corresponding horizontal mean temperature (i.e. $\overline{T} = \tau + \overline{\theta}$) and the magnitudes of the (complex) Fourier amplitudes of the temperature fluctuations (i.e. $|\hat{\theta}_n|$ for $n\neq 0$, the deviations from the horizontal mean) as functions of $z$ from DNS at $Ra = 20000$ in a narrow domain.  The flow consists of a single rising and descending mega-plume and has a similar columnar structure as is observed in wide domains \citep{Hewitt2012, Wen2015JFM}.  In particular, the interior flow is dominated by only a few low-wavenumber Fourier modes, and the Fourier amplitudes $\hat{\theta}_n$ are strongly localized near the walls at high wavenumbers (e.g. $n > 10$).  Given this structure, the spatial/spectral domain can be decomposed into three (or more) subregions at large $Ra$, as schematically shown in figure~\ref{Schematic_DomDecomp}: regions~I and III comprise the near-wall domains in which high resolution in both $x$ and $z$ is required to resolve the small-scale motions in the boundary layers; region~II is the interior region in which coarse resolution in $z$ can be used to capture the relatively large-scale advective motions.  Since the interior flow is controlled by only a few low horizontal-wavenumber Fourier modes (region II(a)), the high-wavenumber subdomain II(b) can be suppressed by setting all field variables to zero there, as we demonstrate subsequently.

The idea underlying the domain decomposition method is simple: at each time step, numerical integrations are performed separately in each subdomain (I, II, III), and then the solutions in each subdomain are patched along their common boundaries (i.e. along $z=Z_1$ and $z=Z_2$) by requiring continuity of the field variables and a finite number of normal derivatives.  To illustrate, we next show how this method can be utilized to solve the momentum equation (\ref{psi}).  An analogous scheme is used to solve the advection--diffusion equation (\ref{theta}) for the temperature fluctuation.

For a given horizontal wavenumber $nk$, (\ref{psi}) becomes
\begin{eqnarray}
    \left[ D^2 - ({n}k)^2 \right]\hat{\psi}_n^{[j]} = -{ink}Ra\hat{\theta}_n^{[j]},\label{psi_discretization}
\end{eqnarray}
where $D$ is the partial derivative operator with respect to $z$, the superscript `$[j]$' denotes the $j$-th subdomain, and both $\hat{\psi}_n$ and $\hat{\theta}_n$ satisfy homogeneous Dirichlet boundary conditions at the heated and cooled walls (i.e. $z = 0$, $1$).  The general solution in the $j$-th subdomain, $\hat{\psi}_n^{[j]}(z)$, can always be written as the sum of a particular integral $\hat{\psi}_{n_p}^{[j]}(z)$ plus two homogeneous solutions, namely
\begin{eqnarray}
    \hat{\psi}_n^{[j]}(z) = \hat{\psi}_{n_p}^{[j]}(z) + \Psi^{[j-1]}\cdot h_L^{[j]}(z) + \Psi^{[j]}\cdot h_U^{[j]}(z),\label{psi_threeparts}
\end{eqnarray}
where $\Psi^{[j-1]}$ and $\Psi^{[j]}$ are the undetermined values of $\hat{\psi}_n$ at the (two) boundaries of the $j$-th subdomain, and the particular integral $\hat{\psi}_{n_p}^{[j]}$ and homogeneous solutions $h_L^{[j]}$ and $h_U^{[j]}$ are defined by
\begin{eqnarray}
    \left[ D^2 - ({n}k)^2 \right]\hat{\psi}_{n_p}^{[j]} = &-{ink}Ra\hat{\theta}_n^{[j]}, \quad &\hat{\psi}_{n_p}^{[j]}(d_{j-1}) = 0, \quad \hat{\psi}_{n_p}^{[j]}(d_{j}) = 0, \label{psi_jp}\\
    \left[ D^2 - ({n}k)^2 \right]h_L^{[j]} = &0, &h_L^{[j]}(d_{j-1}) = 1, \quad h_L^{[j]}(d_{j}) = 0,\label{psi_jhL}\\
    \left[ D^2 - ({n}k)^2 \right]h_U^{[j]} = &0, &h_U^{[j]}(d_{j-1}) = 0, \quad h_U^{[j]}(d_{j}) = 1,\label{psi_jhU}
\end{eqnarray}
where $d_{j}$ denotes the boundary between $j$-th subdomain and ($j$+1)-th subdomain.  
The definitions in (\ref{psi_threeparts})--(\ref{psi_jhU}) imply that continuity of $\hat{\psi}_n$ is automatically enforced at the subdomain boundaries, i.e. $\hat{\psi}_n(z = d_j) = \Psi^{[j]}$.

An obvious advantage of the domain decomposition method is that the grid points can be strategically distributed so that, for example, they can be clustered near the upper and lower walls without simultaneously needing to increase the degrees of freedom used in the domain interior.  Furthermore, a novel and indeed crucial aspect of the proposed domain decomposition scheme is that, in accord with the specific columnar flow structure arising in porous media convection at large $Ra$, region~II(b) can be omitted, with homogeneous Dirichlet boundary conditions enforced directly on the upper and lower interfaces of this subdomain.
A second advantage of domain decomposition is that the uncoupled equations (\ref{psi_jp})--(\ref{psi_jhU}) can be solved \emph{independently} on each subdomain.  However, to obtain the elemental solution $\hat{\psi}_n^{[j]}(z)$ in (\ref{psi_threeparts}), one needs to compute ($\mathcal{M}+1$) unknown $\Psi^{[j]}$, the values of $\hat{\psi}_n$ at the subdomain boundaries, where $\mathcal{M}$ denotes the total number of subdomains in the entire domain (e.g. $\mathcal{M} = 3$ in figure~\ref{Schematic_DomDecomp}).  The terminal values of $\Psi^{[j]}$ can be determined by the boundary conditions at the upper and lower walls: $\Psi^{[0]} = \hat{\psi}_n|_{z=0} = 0$; $\Psi^{[\mathcal{M}]} = \hat{\psi}_n|_{z=1} = 0$.  The remaining ($\mathcal{M}-1$) domain boundary values of $\hat{\psi}_n$ are determined by enforcing continuity of the first derivative across each of the interior interfaces, which gives
\begin{eqnarray}
    h_L^{[j]\prime}\Psi^{[j-1]} + [h_U^{[j]\prime} - h_L^{[j+1]\prime}]\Psi^{[j]} - h_U^{[j+1]\prime}\Psi^{[j+1]} = \hat{\psi}_{n_p}^{[j+1]\prime} - \hat{\psi}_{n_p}^{[j]\prime}, j = 1, \cdots, \mathcal{M}-1, \label{firstderivativecontinuity}
\end{eqnarray}
where the prime denotes the first derivative with respect to $z$, and all derivatives are computed at the interior interface shared by the $j$-th and ($j$+1)-th subdomain ($z = d_j$ and $z=d_{j-1}$, respectively) once the particular integral and homogeneous solutions are obtained by solving (\ref{psi_jp})--(\ref{psi_jhU}).  
A final advantage of the domain decomposition method is that the condition number is also decreased since the size of the matrix algebraic problem in each subdomain is reduced relative to that for a single domain.

\subsection{Hybrid Reduced Model}\label{sec:HybridModel}

As demonstrated in \citep{CDZD2011, Wen2013, WenChini2018JFM}, an \emph{a priori} eigenbasis can be obtained by minimizing the heat-flux functional $nu \equiv \int_0^1(\partial_z\tau)^2dz$ subject to the `spectral constraint'  
\begin{eqnarray}
\min\limits_{{\vartheta}}\left\{{\mathcal{H}}({\vartheta}) \equiv \frac{1}{L}\int_0^1\int_0^L\left(|\nabla{\vartheta}|^2 + w{\vartheta}\partial_z\tau\right)dxdz\right\} \ge 0.\label{EnergyConstraint}
\end{eqnarray}
This energy-stability-like constraint is equivalent to requiring the non-positivity of the ground state eigenvalue $\lambda^0$ of the self-adjoint problem
\begin{eqnarray}
2\nabla^{2}{\vartheta} - \partial_z\tau\,W + \partial_x^{2} \gamma &=& \lambda{\vartheta}, \label{Eig_theta}\\
\nabla^{2}W - Ra\partial_x^{2} {\vartheta}  &=& 0, \label{Eig_w}\\
\nabla^{2}\gamma + Ra\partial_z\tau\,{\vartheta} &=& 0, \label{Eig_Gamma}
\end{eqnarray}
where $\vartheta(x,z)$ is any arbitrary test function satisfying $L$-periodic boundary conditions in $x$ and homogeneous Dirichlet boundary conditions in $z$, and $\gamma(x,z)$ is the Lagrange-multiplier field enforcing the local constraint (\ref{Eig_w}).  Substituting the following
decompositions
\begin{eqnarray}
    {\vartheta}(x, z) &=& \sum\limits_{n=-\infty}^{\infty}\sum\limits_{m=0}^{\infty}\Theta_{mn}(z)e^{inkx}, \label{theta_eigdecomp}\\
    W(x, z) &=& \sum\limits_{n=-\infty}^{\infty}\sum\limits_{m=0}^{\infty}W_{mn}(z)e^{inkx}, \label{W_eigdecomp}\\
    \gamma(x, z) &=& \sum\limits_{n=-\infty}^{\infty}\sum\limits_{m=0}^{\infty}\Gamma_{mn}(z)e^{inkx} \label{gamma_eigdecomp}
\end{eqnarray}
into (\ref{Eig_theta})--(\ref{Eig_Gamma}) and solving the resulting self-adjoint eigenvalue problem,
\begin{eqnarray}
-2\left[D^2 - (nk)^2\right]\Theta_{mn} + \tau'W_{mn} + (nk)^2\Gamma_{mn} &=& \lambda_{mn}\Theta_{mn}, \label{Eig_theta_n}\\
\left[D^2 - (nk)^2\right]W_{mn} + (nk)^2Ra\Theta_{mn} &=& 0, \label{Eig_W_n}\\
\left[D^2 - (nk)^2\right]\Gamma_{mn} + Ra\tau'\Theta_{mn} &=& 0, \label{Eig_Gamma_n}
\end{eqnarray}
for each horizontal wavenumber $nk$ utilizing the two-step algorithm developed in \cite{Wen2013,Wen2015PRE} yields a complete set of orthogonal eigenfunctions,  i.e. $\Theta_{mn}(z)$, $W_{mn}(z)$ and $\Gamma_{mn}(z)$, \emph{and} the background profile $\tau(z)$.

\floatsetup[figure]{style=plain,subcapbesideposition=top}
\begin{figure}[t]
   {\centering
    {\includegraphics[width=0.8\textwidth]{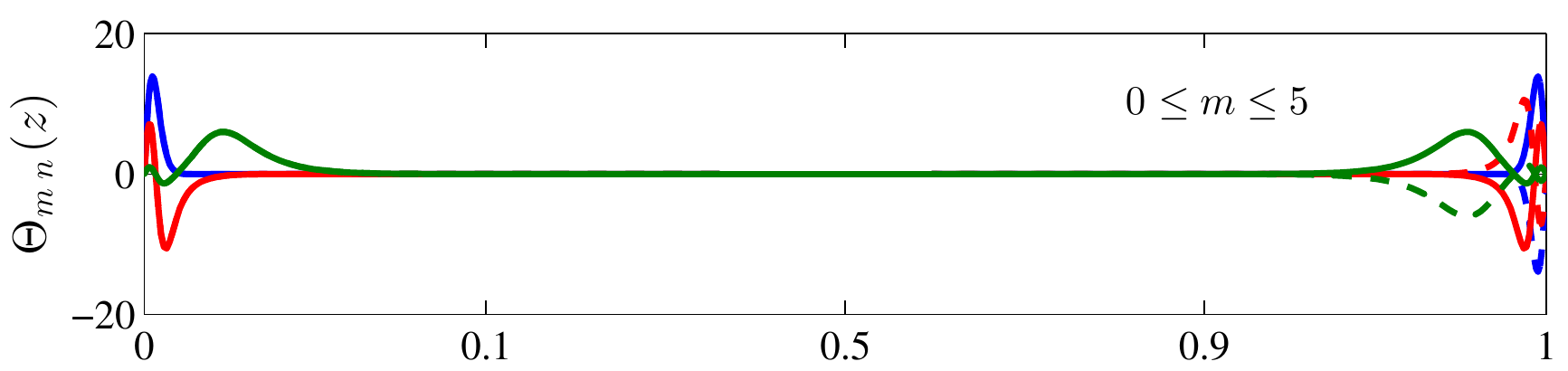}}\\
    {\includegraphics[width=0.8\textwidth]{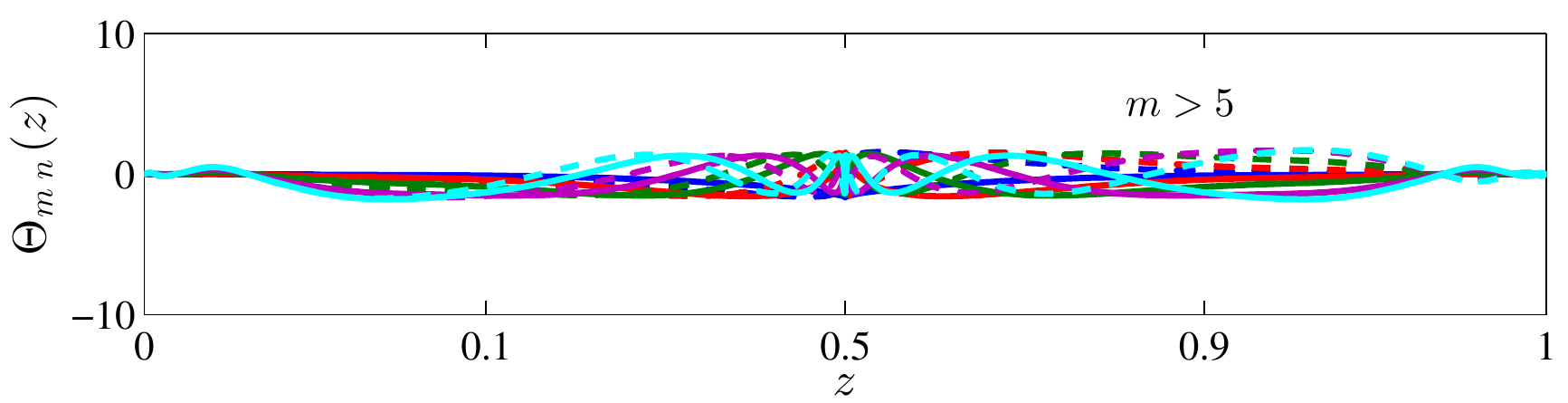}}  \par}
    \caption{Upper bound eigenfunctions for $0 \leq m \leq 15$ at $Ra = 20000$, $L = 0.24$ and $n = 50$.  At large horizontal mode number $n$, the eigenfunctions $\Theta_{mn}(z)$ are localized near the walls for $0 \le m \le M_1$ ($M_1 = 5$ in this case), but exhibit an interior structure for $m > M_1$.  Note that the $z$ values on the horizontal axis are non-uniformly spaced to clearly show the structure of the eigenfunctions.}
    \label{Eigenfunction}
\end{figure}

\begin{figure}[t]
    \centering
    \includegraphics[width=0.6\textwidth]{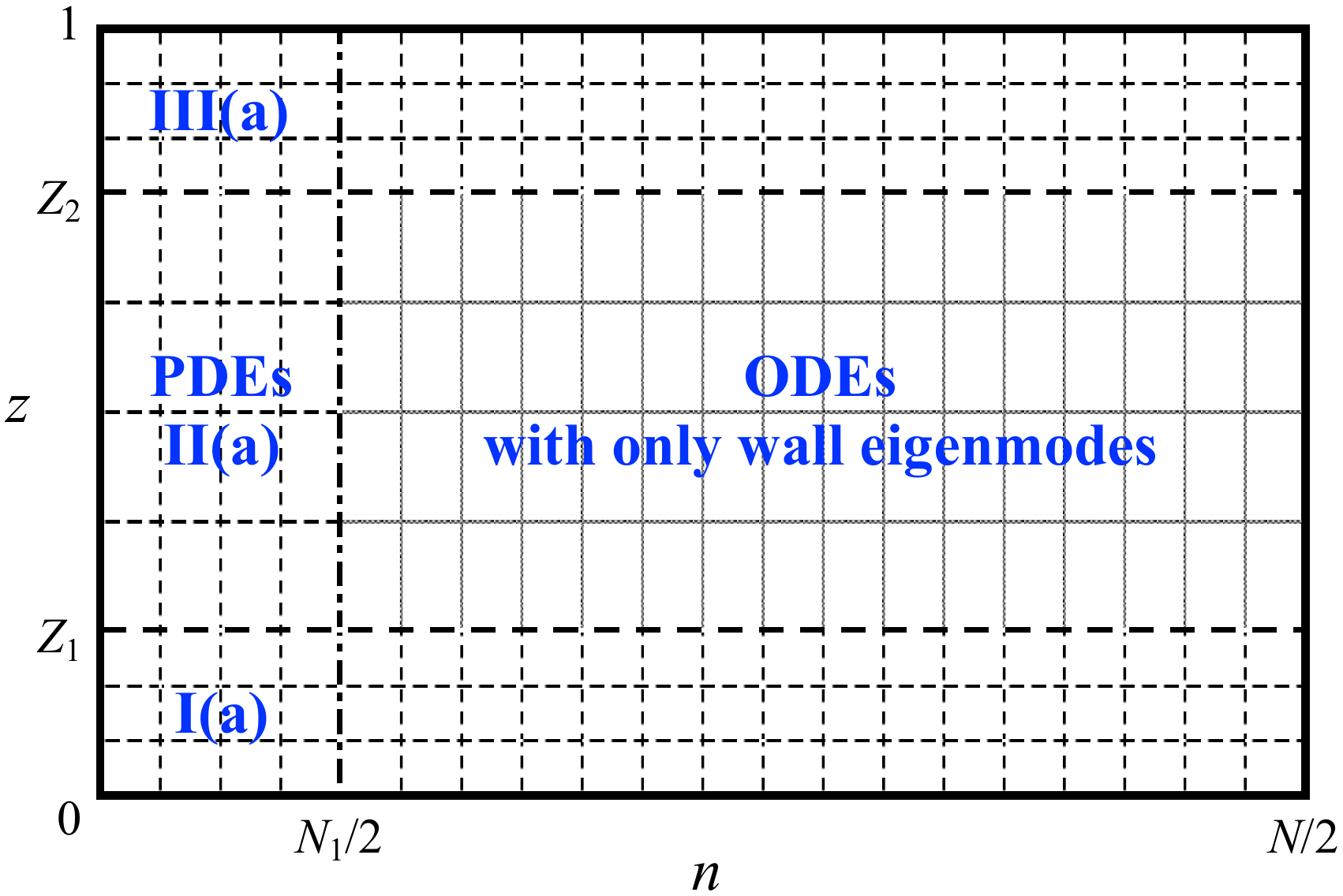}
    \caption{Schematic illustrating hybrid modeling scheme in physical/Fourier space at large $Ra$.  The plot only shows the positive wavenumber regime ($0 \le n \le N/2$), since the Fourier amplitudes are complex conjugates at negative wavenumbers.  For $0 \le n \le N_1/2$, PDEs are solved using the domain decomposition method over subdomains I(a), II(a) and III(a); for $N_1/2 < n \le N/2$, ODEs obtained by projection onto the (upper-bound) wall eigenmodes are solved. Note that $N_1$ in I(a) and III(a) can be larger than that used in II(a).}
    \label{Schematic_Hybrid}
\end{figure}

The upper bound algorithm summarized above thus provides an \emph{a priori} procedure for generating an eigenbasis that is naturally adapted to the dynamics of porous media convection at given parameter values.  Galerkin projection of the governing PDEs onto the upper-bound eigenfunction basis then yields a system of ODEs.  As shown in~figure~\ref{Eigenfunction}, the upper bound eigenfunctions generally exhibit two distinct types of structure at large $Ra$: for small vertical mode number $m$ (e.g. $0 \le m \le 5$), the eigenfunctions are strongly localized near the upper and lower walls, and hereafter are referred to as wall eigenmodes/eigenfuntions; for larger $m$ (e.g. $m > 5$), the eigenfunctions have most of their support in the interior of the domain and, hence, are referred to as interior eigenmodes/eigenfuntions.  This mode separation is more obvious at large horizontal wavenumber.  Recalling that at large $Ra$ the high $x$-wavenumber motions are confined near the upper and lower walls (see~figure~\ref{Tandthetahat_MFU}), it is reasonable to expect that the upper-bound wall eigenmodes will effectively capture the small-scale near-wall flow structures. 
Based on these observations, a hybrid model can be constructed as follows: for small $n$, PDEs are solved to resolve the low- and moderate-wavenumber dynamics, namely, the dynamics of the interior mega-plumes and near-wall proto-plumes; for large $n$, ODEs obtained by projection onto a modest number of wall eigenmodes are solved to compute the small-scale dynamics within the thermal boundary layers.  This hybrid modeling strategy is schematically depicted in~figure~\ref{Schematic_Hybrid}.

In \cite{CDZD2011}, a $6$-mode strictly-ODE model of porous media convection at $Ra=100$ was derived via Galerkin projection onto upper-bound eigenfunctions.  The algorithm for deriving these ODEs (at arbitrary $Ra$) is explained next.

First, all dependent field variables are decomposed according to
\begin{eqnarray}
    \theta(x, z, t) &=& \sum\limits_{n=-\infty}^{\infty}\sum\limits_{m=0}^{\infty}{a_{mn}(t)}\Theta_{mn}(z)e^{inkx}, \label{theta_decomposition}\\
    w(x, z, t) &=& \sum\limits_{n=-\infty}^{\infty}\sum\limits_{m=0}^{\infty}{b_{mn}(t)}W_{mn}(z)e^{inkx}, \label{w_decomposition}\\
    u(x, z, t) &=& \sum\limits_{n=-\infty}^{\infty}\sum\limits_{m=0}^{\infty}{c_{mn}(t)}U_{mn}(z)e^{inkx}, \label{u_decomposition}
\end{eqnarray}
where $a_{mn}$, $b_{mn}$ and $c_{mn}$ are, respectively, the time-dependent coefficients of the temperature fluctuation about $\tau(z)$ (i.e. not about the horizontal mean temperature $\overline{T}(z,t)$), the vertical velocity component and the horizontal velocity component.  The vertical eigenfunctions $\Theta_{mn}(z)$ and $W_{mn}(z)$ satisfy (\ref{Eig_theta_n})--(\ref{Eig_Gamma_n}) subject to the following orthonormalization condition:
\begin{eqnarray}
    \int_0^1 \Theta_{mn}(z)\Theta_{pn}(z)dz = \delta_{mp}, \label{orthonormalization}
\end{eqnarray}
with the unit delta function $\delta_{mp} = 1$ for $m = p$ and $\delta_{mp} = 0$ for $m \ne p$ (for integer $p$), and
\begin{eqnarray}
    c_{mn}(t)U_{mn}(z) = \frac{-1}{ink}b_{mn}(t)DW_{mn}(z) \label{ucoef}
\end{eqnarray}
for $n \ne 0$ using the divergence-free condition on the velocity field.  Moreover, (\ref{Eig_w}) also implies that $b_{mn}(t) = a_{mn}(t)$ for $n \ne 0$.

Substituting (\ref{theta_decomposition})--(\ref{u_decomposition}) into the temperature fluctuation equation (\ref{theta}) and forming the inner product of each term with $\Theta_{pr}(z)\text{exp}(irkx)$, for integer $r$, yields a system of ODEs for $n \ne 0$:
\begin{eqnarray}
    \dot{a}_{mn} &=& \mu_{mn}a_{mn} + \sum_{\underset{(p\ne m)}{p=0}}^{\infty}\mu_{mn}^p a_{pn} +
    \sum_{\underset{(j\ne n)}{j=-\infty}}^{\infty}\sum_{p=0}^{\infty}\sum_{q=0}^{\infty}
        \Lambda_{mn}^{jpq}a_{qj}a_{p(n-j)}, \label{ODE_n_notzero}
\end{eqnarray}
where the linear coefficients 
\begin{eqnarray}
    \mu_{mn} &=& -\lambda_{mn}/2 \leq 0 \quad (\mbox{where}\,\lambda_{mn}\;\text{is obtained from (\ref{Eig_theta_n})}), \label{mu_mn}\\
    \mu_{mn}^p &=&  \left(\frac{n^2k^2}{2}\right)\int_0^1 \left[ \Theta_{mn}\Gamma_{pn} - \Theta_{pn}\Gamma_{mn} \right] dz, \label{mu_mnp}
\end{eqnarray}
and the nonlinear coefficients
\begin{eqnarray}
    \Lambda_{mn}^{jpq} = \int_0^1\Theta_{mn}\left[ \left( \frac{j}{n-j}\right) DW_{p(n-j)}\Theta_{qj} - W_{p(n-j)}D\Theta_{qj} \right]dz. \label{nonlinear_coef}
\end{eqnarray}
For $n = 0$, the eigenfunctions can be obtained analytically from (\ref{Eig_theta_n})--(\ref{Eig_Gamma_n}):
\begin{eqnarray}
   \Theta_{m0}(z) = \sqrt{2}\sin\left((m+1)\pi z\right); \quad W_{m0}(z) = U_{m0}(z) = 0. \label{Eigenfunctions_mean}
\end{eqnarray}
Galerkin projection of the horizontal average of (\ref{theta}) onto $\Theta_{p0}(z)$ then yields the amplitude equations for $a_{m0}(t)$:
\begin{eqnarray}
\dot{a}_{m0} = \mu_{m0}a_{m0} + \sum_{\underset{(j\ne 0)}{j=-\infty}}^{\infty}\sum_{p=0}^{\infty}\sum_{q=0}^{\infty}
                \Lambda_{m0}^{jpq}a_{qj}a_{pj}^* + f_m, \label{ODE_n_zero}
\end{eqnarray}
where the asterisk denotes complex conjugation and the coefficients satisfy
\begin{eqnarray}
    \mu_{m0} &=& -(m+1)^2\pi^2, \label{mu_m0}\\
    \Lambda_{m0}^{jpq} &=& \sqrt{2}(m+1)\pi \int_0^1 W_{pj}\Theta_{qj}\cos((m+1)\pi z)dz, \label{nonlinear_coef_n0}\\
    f_m &=& \sqrt{2}(m+1)\pi \left[1 - (m+1)\pi \int_0^1\tau(z)\sin((m+1)\pi z)dz\right]. \label{fm}
\end{eqnarray}
Finally, low-dimensional models are obtained by suitably truncating the sums in (\ref{ODE_n_notzero}) and (\ref{ODE_n_zero}).

At large $Ra$, solving a large system of ODEs for the amplitudes $a_{mn}(t)$ is inefficient.  Unfortunately, ODE models obtained by low-order truncations prove insufficiently accurate.  Thus, we propose a hybrid modeling scheme in which the governing PDEs are solved for $0 \le n \le N_1/2$ using a Fourier--Chebyshev collocation method, while the subset of ODEs in (\ref{ODE_n_notzero}) generated by projection strictly onto wall eigenfunctions is employed for $n > N_1/2$ to model the high-wavenumber near-wall motions.  Hence, for $n > N_1/2$, the Fourier components of the temperature field can be approximated as
\begin{eqnarray}
    \hat{\theta}_{n}(z,t) \approx \sum\limits_{m=0}^{M_1}{a_{mn}(t)\Theta_{mn}(z)}, \label{theta_wallmode}
\end{eqnarray}
where each of the eigenfunctions $\Theta_{mn}(z)$ for $0 \le m \le M_1$ is wall-localized (see~figure~\ref{Eigenfunction}).  The efficiency of the required ODE computations can be dramatically increased by avoiding direct summation of the nonlinear convolutions in (\ref{ODE_n_notzero}).  To do this, recall that the nonlinear terms in the ODE governing the evolution of mode $a_{mn}$ are obtained by projecting the nonlinear terms in the governing PDEs onto the upper bound eigenfunctions $\Theta_{mn}$, \emph{viz.}
\begin{eqnarray}
    \mathcal{N}_{mn}^{\text{ODE}} \equiv \sum_{\underset{(j\ne n)}{j=-\infty}}^{\infty}\sum_{p=0}^{\infty}\sum_{q=0}^{\infty}
        \Lambda_{mn}^{jpq}a_{qj}a_{p(n-j)} = \int_0^1\widehat{\mathcal{N}}_{n}^{\text{PDE}} \Theta_{mn}dz, \label{nonlinear_projection}
\end{eqnarray}
where $\widehat{\mathcal{N}}_{n}^{\text{PDE}}$ denotes the Fourier components of the nonlinear terms $\left(-u\partial_x\theta - w\partial_z\theta\right)$ for a given horizontal wavenumber $nk$.  Since the Fast Fourier Transform (FFT) algorithm can be used to efficiently compute $\widehat{\mathcal{N}}_{n}^{\text{PDE}}$, the nonlinear terms $\mathcal{N}_{mn}^{\text{ODE}}$ are determined here according to the following steps.  Suppose the coefficients $a_{mn}$ and $b_{mn}$ are given initially. First, the Fourier components of the temperature fluctuation and velocity fields are computed, respectively, through (\ref{theta_wallmode}) and
\begin{eqnarray}
    \hat{u}_n &=& \left(\frac{-1}{ink}\right)\sum\limits_{m=0}^{M_1}{b_{mn}DW_{mn}}, \label{u_hat_n_ODE}\\
    \hat{w}_n &=& \sum\limits_{m=0}^{M_1}{b_{mn}W_{mn}}. \label{w_hat_n_ODE}
\end{eqnarray}
$\widehat{\mathcal{N}}_{n}^{\text{PDE}}$ is then computed using the standard pseudo-spectral method for solving PDEs.  Finally, $\mathcal{N}_{mn}^{\text{ODE}}$ is obtained by projecting $\widehat{\mathcal{N}}_{n}^{\text{PDE}}$ onto $\Theta_{mn}$ using (\ref{nonlinear_projection}).  Compared with the pseudospectral numerical solution of the governing PDEs, there exist two obvious advantages of solving ODEs: first, it is not necessary to solve the momentum equation (\ref{psi}) each time step once the eigenfunctions are obtained (i.e. in a \emph{pre-}processing step); secondly, since only $4$--$6$ wall eigenfunctions are retained for each horizontal mode $n$ (for $n>N_1/2$), the total number of modes is significantly reduced.  Although the projection step (\ref{nonlinear_projection}) incurs modest additional computations, nevertheless the total operation count is significantly decreased since the number of ODE modes used for each $n$ is so small (e.g. $M_1=5$).

For a given horizontal mode $n$, these ODEs can be written in matrix form
\begin{eqnarray}
    \left[ \begin{array}{c} \dot{a}_{0n} \\ \dot{a}_{1n} \\ \dot{a}_{2n} \\ \dot{a}_{3n} \\ \vdots \end{array} \right] =
    \left[ \begin{array}{ccccc} \mu_{0n} & \mu_{0n}^{1} & \mu_{0n}^{2} & \mu_{0n}^{3} & \cdots \\
                                \mu_{1n}^0 & \mu_{1n}   & \mu_{1n}^{2} & \mu_{1n}^{3} & \cdots \\
                                \mu_{2n}^0 & \mu_{2n}^1 & \mu_{2n}     & \mu_{2n}^3   & \cdots \\
                                \mu_{3n}^0 & \mu_{3n}^1 & \mu_{3n}^2   & \mu_{3n}     & \cdots \\
                                \vdots     & \vdots     &              & \ddots       &          \end{array} \right]
    \left[ \begin{array}{c} a_{0n} \\ a_{1n} \\ a_{2n} \\ a_{3n} \\ \vdots \end{array} \right] +
    \left[ \begin{array}{c} \mathcal{N}_{0n} \\ \mathcal{N}_{1n} \\ \mathcal{N}_{2n} \\ \mathcal{N}_{3n} \\ \vdots \end{array} \right]^{\text{ODE}}, \label{ODE_matrixform}
\end{eqnarray}
or in vector form
\begin{eqnarray}
    \dot{\vec{a}}_n = \mathcal{A}_n\vec{a}_n + \vec{\mathcal{N}}_n^{\text{ODE}}, \label{ODE_matrixform_simple}
\end{eqnarray}
where $\mathcal{A}_n$ is the matrix of linear coefficients and $\vec{\mathcal{N}}_n$ is the vector of nonlinear terms.

\section{Results and Discussion}\label{sec:Results}
\floatsetup[figure]{style=plain,subcapbesideposition=top}  
\begin{figure}[t!]
    \centering
    \sidesubfloat[]{\includegraphics[height=2.7in]{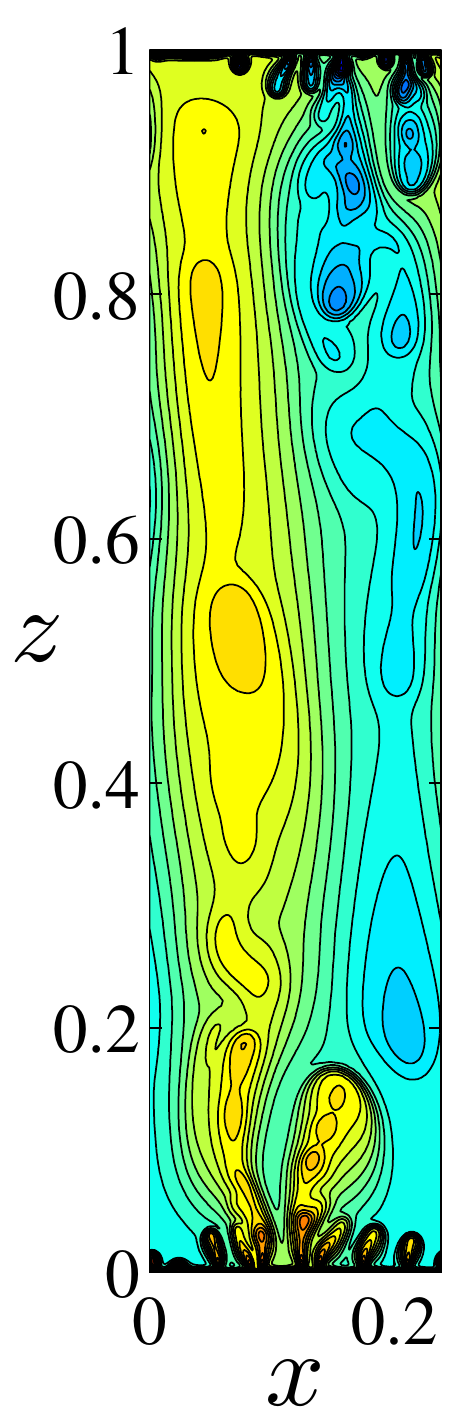}} \quad\quad
    \sidesubfloat[]{\includegraphics[height=2.7in]{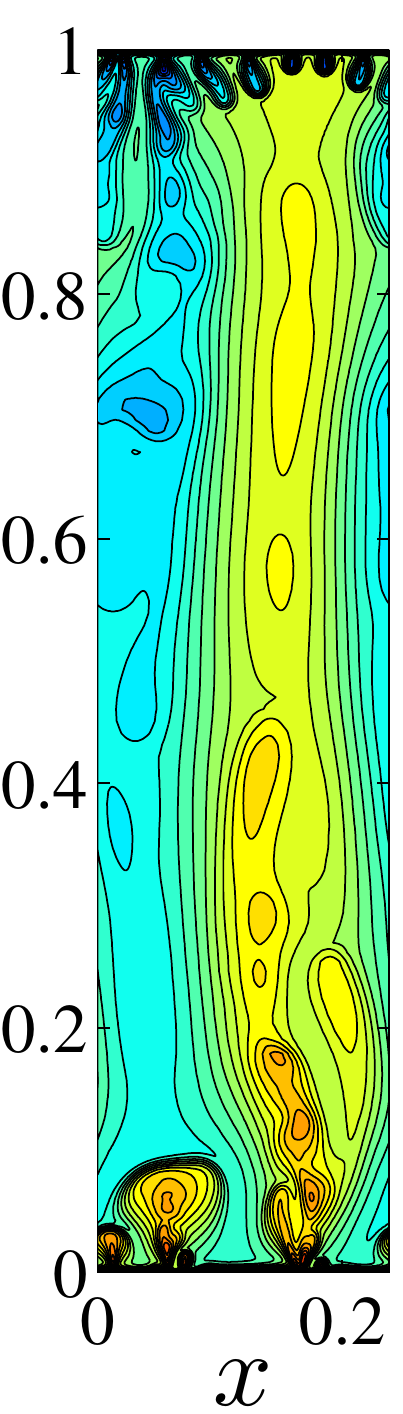}} \quad\quad
    \sidesubfloat[]{\includegraphics[height=2.7in]{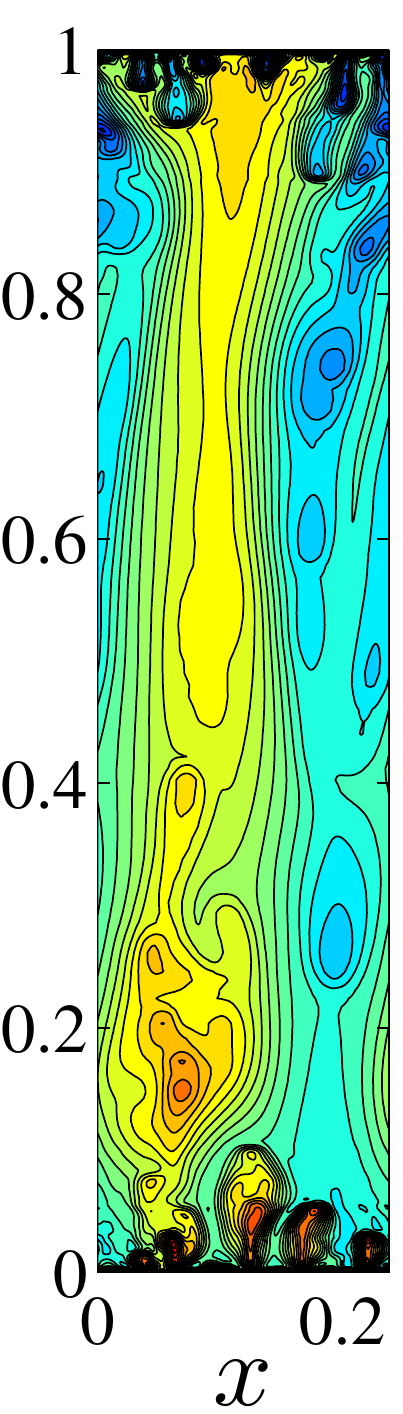}} \quad\quad
    \sidesubfloat[]{\includegraphics[height=2.7in]{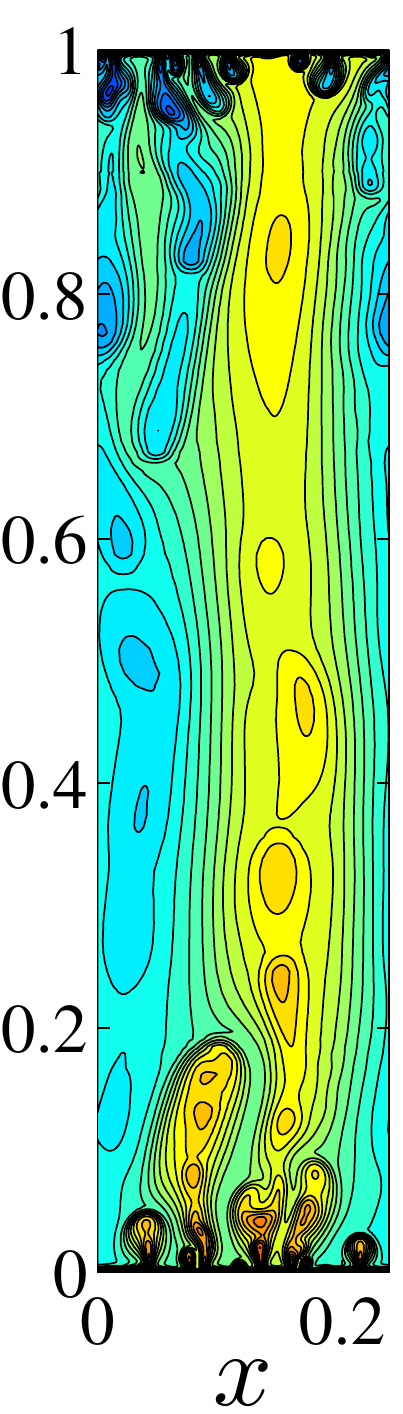}} \par
    \caption{Snapshots of the temperature fields from ($a$) DNS, ($b$) domain decomposition method, ($c$) hybrid model 1 and ($d$) hybrid model 2 at $Ra = 20000$ and $L = 0.24$.  (See table~\ref{ResolutionNu} for further details of these simulations.)}
    \label{Snapshot_3methods}
\end{figure}

In this study, $L = 4\pi Ra^{-0.4}$ is utilized as the approximate width of the minimal flow unit.  Temporal discretization of both the PDEs and ODEs is achieved using the Crank--Nicolson method for the linear terms and a two-step Adams--Bashforth method for the nonlinear terms, yielding second-order accuracy in time.  In the computations employing the domain decomposition method, the domain is split into three regions with $Z_1 = 0.1$ and $Z_2 = 0.9$; $N_1/2 = 10$ in the interior region II(a); and the interior high-wavenumber region II(b) is omitted.  Although fixed in all of the simulations reported here, $Z_1=1-Z_2$ may be expected to vary roughly as $Ra^{-0.4}$ to track the vertical scale of the near-wall proto-plumes.  Moreover, two hybrid models (labeled 1 and 2) are developed: for hybrid model~1, $N_1/2 = 10$ is used in the near-wall subdomains I(a) and III(a); and for hybrid model~2, $N_1 \approx N/3$ in those subdomains for improved accuracy.

Figure~\ref{Snapshot_3methods} shows snapshots of the temperature fields obtained from DNS, the domain decomposition method and the hybrid models in the minimal flow unit at $Ra = 20000$.  Both the domain decomposition method and the hybrid models capture the primary structural features of the turbulent columnar flow, with well organized mega-plumes in the interior and more complex flow structures near the heated and cooled walls.  Table~\ref{ResolutionNu} documents the resolution, Nusselt number and efficiency for the four simulations.  Remarkably, compared with the DNS, the relative error in $Nu$ obtained using the domain decomposition method is less than $1\%$, confirming that an emergent dynamical feature of high-$Ra$ porous media convection is the suppression of the interior high-wavenumber region (II(b) in~figure~\ref{Schematic_DomDecomp}). 

\begin{table}[t!]
{\centering
\begin{tabular}{|l|c|c|c|c|c|c|c|}
\hline
\multirow{2}*{Method} & \multirow{2}*{$\dfrac{N_1}{2}$} & \multirow{2}*{$\dfrac{N}{2}$} & \multirow{2}*{$M$} & \multirow{2}*{$\#$ Modes} & \multirow{2}*{$Nu$} & \multirow{2}*{Error} & \multirow{1}*{CPU time (s)}\\
       &       &       &       &       &       &       & \multirow{1}*{for $10^3$ steps}\\
\hline
\multirow{2}*{DNS} & \multirow{2}*{128} & \multirow{2}*{128} & \multirow{2}*{320} & \multirow{2}*{82176} & \multirow{2}*{139.7} & \multirow{2}*{0} & \multirow{2}*{1465}\\
       &       &       &       &       &       &       & \\
\hline
\multirow{2}*{UR-DNS} & \multirow{2}*{40} & \multirow{2}*{40} & \multirow{2}*{320} & \multirow{2}*{25680} & \multirow{2}*{213.1} & \multirow{2}*{52.5\%} & \multirow{2}*{506}\\
       &       &       &       &       &       &       & \\
\hline
\multirow{2}*{DD I} & \multirow{2}*{10} & \multirow{2}*{128} & \multirow{2}*{64} & \multirow{2}*{16640} & \multirow{6}*{138.9} & \multirow{6}*{0.57\%} & \multirow{6}*{231} \\
       &       &       &       &       &       &       & \\
\multirow{2}*{DD II(a)} & \multirow{2}*{10} & \multirow{2}*{10} & \multirow{2}*{128} & \multirow{2}*{2580} &  &  & \\
       &       &       &       &       &       &      & \\
\multirow{2}*{DD III} & \multirow{2}*{10} & \multirow{2}*{128} & \multirow{2}*{64} & \multirow{2}*{16640} & \multirow{2}*{} & \multirow{2}*{} & \\
       &       &       &       &       &       &      & \\
\hline
\multirow{6}*{HM 1} & \multirow{2}*{10} & \multirow{2}*{128} & \multirow{2}*{PDE: I(a) 64,} & \multirow{3}*{PDE: 5180} & \multirow{6}*{182.5} & \multirow{6}*{30.6\%} & \multirow{6}*{109} \\
       &       &       &       &       &       &      & \\
       & \multirow{2}*{10} & \multirow{2}*{10} &   \multirow{2}*{II(a) 128,}    &       &       &      & \\
       &       &       &       & \multirow{3}*{ODE: 1416} &       &      & \\
       & \multirow{2}*{10} & \multirow{2}*{128} & \multirow{2}*{III(a) 64; ODE:5} &      &       &      & \\
       &       &       &       &      &       &      & \\
\hline
\multirow{6}*{HM 2} & \multirow{2}*{40} & \multirow{2}*{128} & \multirow{2}*{PDE: I(a) 64,} & \multirow{3}*{PDE: 12980} & \multirow{6}*{168.6} & \multirow{6}*{20.7\%} & \multirow{6}*{173} \\
       &       &       &       &       &       &      & \\
       & \multirow{2}*{10} & \multirow{2}*{10} & \multirow{2}*{II(a) 128,} &       &       &      & \\
       &       &       &       & \multirow{3}*{ODE: 1056}  &       &      & \\
       & \multirow{2}*{40} & \multirow{2}*{128} & \multirow{2}*{III(a) 64; ODE:5} &  &       &      & \\
       &       &       &       &       &       &      & \\
\hline
\end{tabular}
\par}
\caption{Resolution, Nusselt number and efficiency for various computational schemes at $Ra = 20000$ and $L = 0.24$.  All simulations are performed using a single thread in Matlab.  Results obtained from the fully-resolved DNS are treated as being exact for purposes of computing the error in the Nusselt number obtained with the other methods. For reference, results from an under-resolved DNS (UR-DNS) with high wavenumbers removed are included in the third row. DD: domain decomposition method.  HM: hybrid model.   Hybrid model~1, for which less than one-tenth the number of modes used in the DNS is retained, nevertheless yields an estimate of the Nusselt number with a relative error of only 30\%, yet runs 13 times faster than the DNS. } 
\label{ResolutionNu}
\end{table}

\begin{figure}[h!]
    \centering
    \includegraphics[width=0.7\textwidth]{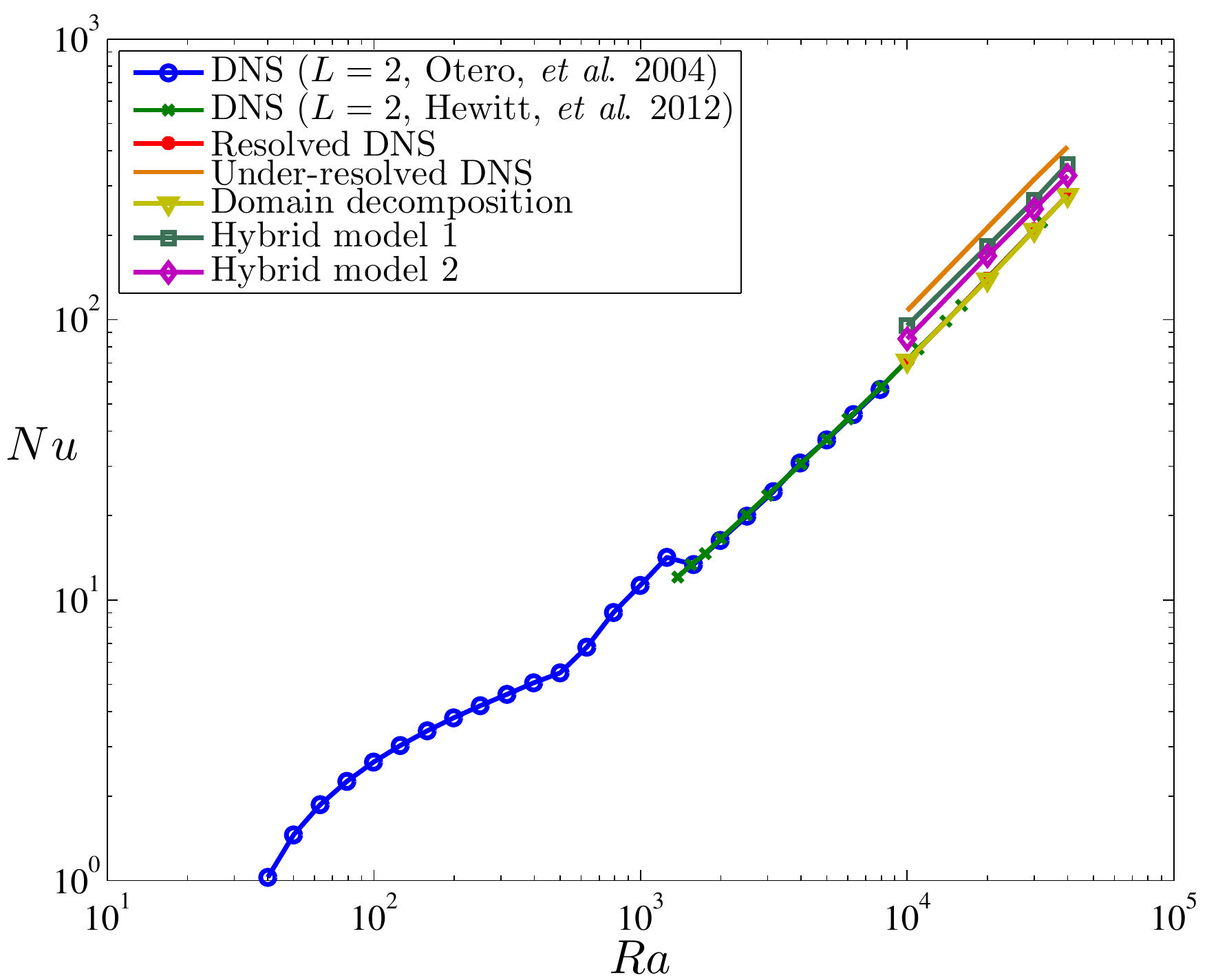}
    \caption{$Nu$ as a function of $Ra$ obtained using various computing strategies at large values of the Rayleigh number.  All the simulations used to generate the results shown here were performed in the minimal flow unit $L = 4\pi Ra^{-0.4}$ except for the studies of \cite{Otero2004} and \cite{Hewitt2012}.  In the domain decomposition and hybrid models, the high-wavenumber interior region II(b) ($n > 10$) was suppressed for each $Ra$; moreover, in the hybrid models, $6$ ODEs for each horizontal mode number $n$ were employed to capture the small-scale dynamics near the walls.}
    \label{NuvsRa_MFU}
\end{figure}

As discussed in previous sections, at large horizontal mode number (e.g. $n > 40$ at $Ra=20000$ and $L=0.24$), the Fourier amplitudes of the temperature fluctuations are strongly localized near the upper and lower walls, where they superpose to comprise the small rolls within the thermal boundary layers.  To investigate the effects of the small-scale motions on heat transport, results computed using an under-resolved DNS, with only low and moderate wavenumbers retained (e.g. $n \le 40$), are also included in table~\ref{ResolutionNu}.  Evidently, the heat flux $Nu$ is (erroneously) \emph{increased} by over $50\%$ due to the exclusion of the small-scale motions within the thermal boundary layers.  For hybrid model~2, however, for which a very small number of upper-bound wall eigenfunctions (ODEs) is utilized to model the thermal boundary-layer dynamics, this error is reduced to about $20\%$ while enabling an 8-fold speed-up relative to DNS.  A 13-fold reduction in CPU time can be achieved using hybrid model~1, which employs only 11 PDEs (including the horizontal-mean equation) to resolve the very low-wavenumber large-scale motions across the entire flow domain; the resulting relative error in $Nu$ is only $30\%$.  It should be noted that in both hybrid models, the modes used in the solution of the PDEs constitute a large fraction of the total modes.  Therefore, future model reduction efforts should target limiting the computational expense of solving these low-wavenumber PDEs.

Figure~\ref{NuvsRa_MFU} shows $Nu$ as a function of $Ra$ as computed from the various algorithms.  The data confirm that convection in a minimal flow unit produces the same heat transport per unit area as is realized in wider domains.  The domain decomposition method with the interior high-wavenumber region suppressed is seen to accurately reproduce the exact Nusselt number for a range of large $Ra$ values.  Furthermore, the upper-bound wall eigenmodes evidently can efficiently represent the small-scale motions within the thermal boundary layers, which, via comparison of the resolved and under-resolved DNS, clearly play a significant role in heat transport in high-$Ra$ porous media convection.  Nevertheless, the discrepancies in $Nu$ as computed using the hybrid models and the DNS (and quantified in table~\ref{ResolutionNu} and figure~\ref{NuvsRa_MFU}) indicate that the upper-bound wall eigenmodes cannot \emph{fully} resolve the small-scale near-wall motions.  Indeed, as shown in figure~\ref{Eigenfunction}, although the interior eigenmodes have most of their support in the domain interior, they do not vanish identically in the boundary-layer regions.  This implies that the upper-bound wall eigenfuntions do not constitute a complete basis with which to fully represent the small-scale near-wall dynamics.

\begin{figure}[t]
   {\centering
    \includegraphics[width=0.7\textwidth]{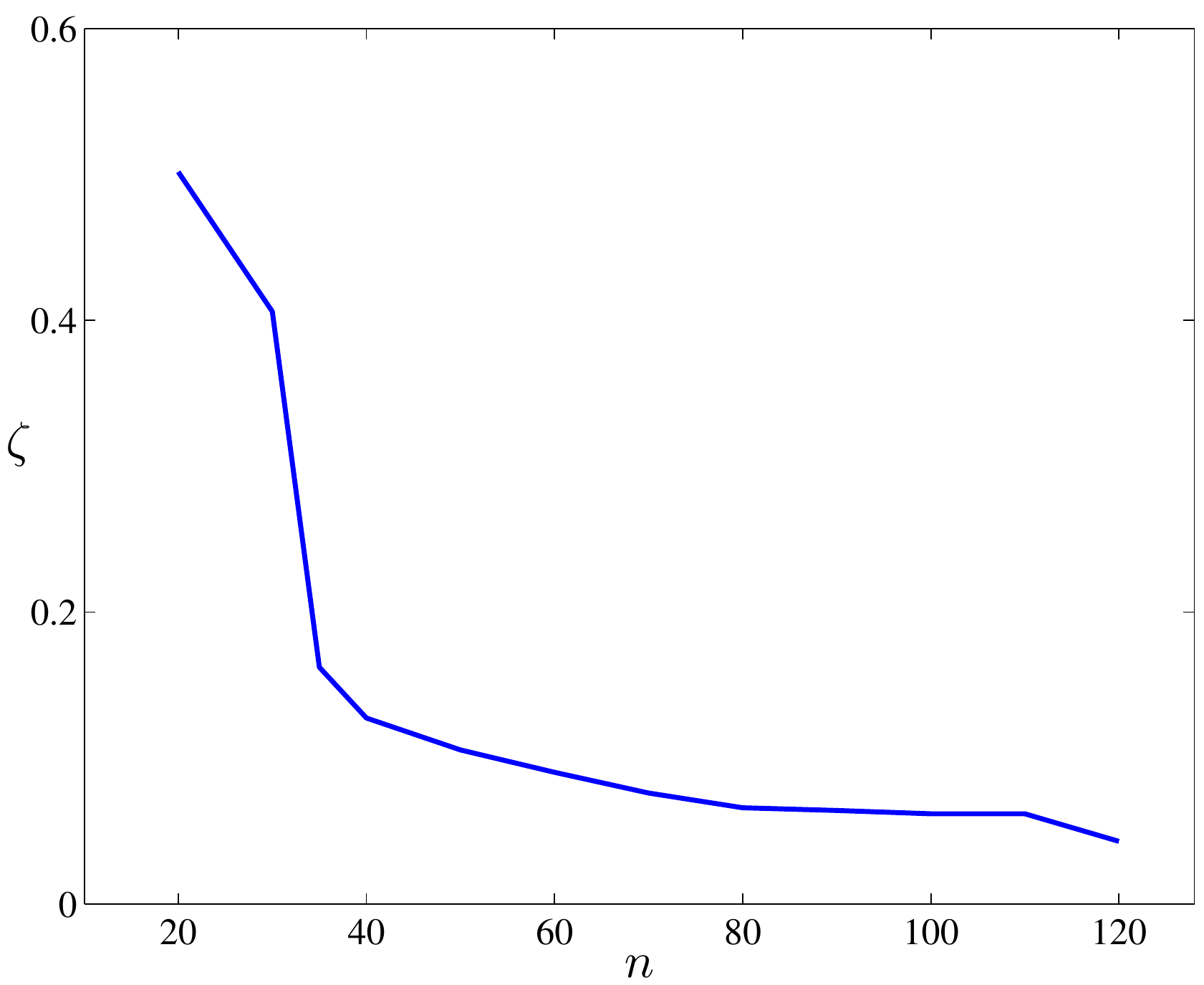}}
    \caption{Time-averaged relative error ($\zeta$) of the projection of $\hat{\theta}_n$ from DNS onto the wall eigenmodes as a function of Fourier mode number $n$ at $Ra = 20000$ and $L=0.24$.  $\zeta$ is defined in~(\ref{Projection_Wall}).}
    \label{Projection}
\end{figure}

To quantitatively assess the suitability of these wall eigenmodes, we perform the following projection:
\begin{eqnarray}
    \hat{\theta}_n^{\text{P}} = \left(\int_0^1\hat{\theta}_n^{\text{DNS}}\Theta_{0n}dz\right)\cdot\Theta_{0n} + \left(\int_0^1\hat{\theta}_n^{\text{DNS}}\Theta_{1n}dz\right)\cdot\Theta_{1n} + \cdots, \label{projection}
\end{eqnarray}
where the DNS data $\hat{\theta}_n^{\text{DNS}}$ are projected only onto the wall eigenmodes at a given horizontal wavenumber $nk$.  If the time-averaged relative error of the projection
\begin{eqnarray}
    \zeta = \bigg{\langle} \frac{||\hat{\theta}_n^{\text{DNS}} - \hat{\theta}_n^{\text{P}}||}{||\hat{\theta}_n^{\text{DNS}}||} \bigg{\rangle}
    \label{Projection_Wall}
\end{eqnarray}
is sufficiently small at each $n$, where $\left\|\cdot\right\|$ denotes a 2-norm and $\langle (\cdot)\rangle$ denotes temporal averaging, then the upper-bound wall eigenfunctions should be able to compactly and accurately represent the dynamics.  Figure~\ref{Projection} shows the variation of $\zeta$ with mode number $n$.  Clearly, the wall eigenfunctions are effective surrogates at large $n$ (high wavenumbers): for $n \ge 40$, instead of using $321$ Chebyshev modes at each $n$ in DNS at $Ra=20000$, $6$ wall eigenmodes can accurately represent more than 80\% of the structure of the small-scale roll motions near the wall, consistent with the error of hybrid model~2 shown in table~\ref{ResolutionNu} (i.e. 20.7\%).  Clearly, however, the wall eigenfunctions cannot adequately capture the low-wavenumber dynamics; consequently, a sufficient number of PDEs must be retained in the hybrid models to achieve reasonable accuracy.

\section{Conclusions}
\label{sec:Conclusions}
In this investigation, two different strategies have been presented to reduce the degrees of freedom in numerical simulations of high-$Ra$ porous media convection: a domain decomposition method and a hybrid model using an eigenbasis from upper bound theory.  In the first strategy, the computational domain is decomposed into three subdomains according to the specific emergent structure of the flow at large $Ra$.  Although only a few low wavenumbers are retained in  the interior, the dynamics is still very well resolved.  In the second strategy, a set of \emph{a priori} eigenfunctions obtained from upper bound analysis of the system is used to model the small-scale motions within the upper and lower thermal boundary layers.  Since these eigenfunctions are directly extracted from the governing equations, they inherent important structural attributes of the flow at the given parameter values.  The results indicate that the upper-bound wall eigenmodes can efficiently and reasonably accurately represent the small-scale dynamics near the upper and lower walls.  
Further reduction in the dimension of the hybrid model is limited by the need to solve a system of PDEs to resolve the low horizontal-wavenumber dynamics coupling the interior mega-plumes with the near-wall proto-plumes.  Future model reduction efforts therefore should seek to reduce the computational expense of solving these low-wavenumber PDEs and, if feasible, to exploit temporal scale separation between the small- and large-scale dynamics.  Nevertheless, relative to DNS, the present hybrid model enables over an order-of-magnitude reduction in CPU time with only a modest loss in accuracy.
Finally, we emphasize that the model reduction strategies proposed in this investigation should yield even more significant computational savings for three-dimensional (3D) simulations of porous media convection, since in 3D the flow retains a similar columnar structure \cite{Pau2010, Fu2013, Hewitt2014} but, crucially, the low-wavenumber modes constitute a smaller fraction of the total number of horizontal modes.



 
\bibliographystyle{model1-num-names}
\bibliography{WC_ReducedModel,WHref,Bibliography}

\end{document}